\documentclass[useAMS,usenatbib,usegraphicx]{mn2e}

\title[Discovery of two magnetic massive stars in the ONC]{Discovery of two magnetic massive stars in the Orion Nebula Cluster: a clue to the origin of neutron star magnetic fields?}
\author[V. Petit et al.]{V. Petit$^{1}$\thanks{E-mail: Veronique.Petit.1@ulaval.ca}, 
G.A. Wade$^{2}$,
L. Drissen$^{1}$,
T. Montmerle$^{3}$ and
E. Alecian$^{2,4}$\thanks{Based on observations obtained at the Canada-France-Hawaii Telescope (CFHT) which is operated by the National Research Concil of Canada, the Institut National des Sciences de l'Univers of the Centre National de la Recherche Scientifique of France, and the University of Hawaii.}\\
$^{1}$D\'ept. de physique, g\'enie physique et optique, Universit\'e Laval, Qu\'ebec (QC), Canada, G1K 7P4 and \\ Centre de recherche en astrophysique du Qu\'ebec\\
$^{2}$Dept. of Physics, Royal Military College of Canada, PO Box 17000, Stn Forces, Kingston, Canada, K7K 4B4\\ 
$^{3}$Laboratoire d'Astrophysique de Grenoble, Universit\'e Joseph Fourier,  CNRS, BP 53, 38041 Grenoble Cedex, France\\
$^{4}$Observatoire de Paris, LESIA, Place Jules Janssen, F-92195 Meudon Cedex, France} 

\begin{document}
%
%  These Macros are taken from the AAS TeX macro package version 4.0.
%  Include this file in your LaTeX source only if you are not using
%  the AAS TeX macro package and need to resolve the macro definitions
%  in the BibTeX entries returned by the ADS abstract service.
%
%  For more information on the AASTeX macro package, please see the URL
%	http://www.aas.org/publications/aastex.html
%  For more information about ADS abstract server, please see the URL
%	http://adswww.harvard.edu/ads_abstracts.html
%

% Abbreviations for journals.  The object here is to provide authors
% with convenient shorthands for the most "popular" (often-cited)
% journals; the author can use these markup tags without being concerned
% about the exact form of the journal abbreviation, or its formatting.
% It is up to the keeper of the macros to make sure the macros expand
% to the proper text.  If macro package writers agree to all use the
% same TeX command name, authors only have to remember one thing, and
% the style file will take care of editorial preferences.  This also
% applies when a single journal decides to revamp its abbreviating
% scheme, as happened with the ApJ (Abt 1991).

\def\jnl@style{\it}
%commente par Seb
\def\aaref@jnl#1{{\jnl@style#1}}
%ref remplace par aaref pour eviter conflit...

\def\aaref@jnl#1{{\jnl@style#1}}

\def\aj{\aaref@jnl{AJ}}                   % Astronomical Journal
\def\araa{\aaref@jnl{ARA\&A}}             % Annual Review of Astron and Astrophys
\def\apj{\aaref@jnl{ApJ}}                 % Astrophysical Journal
\def\apjl{\aaref@jnl{ApJ}}                % Astrophysical Journal, Letters
\def\apjs{\aaref@jnl{ApJS}}               % Astrophysical Journal, Supplement
\def\ao{\aaref@jnl{Appl.~Opt.}}           % Applied Optics
\def\apss{\aaref@jnl{Ap\&SS}}             % Astrophysics and Space Science
\def\aap{\aaref@jnl{A\&A}}                % Astronomy and Astrophysics
\def\aapr{\aaref@jnl{A\&A~Rev.}}          % Astronomy and Astrophysics Reviews
\def\aaps{\aaref@jnl{A\&AS}}              % Astronomy and Astrophysics, Supplement
\def\azh{\aaref@jnl{AZh}}                 % Astronomicheskii Zhurnal
\def\baas{\aaref@jnl{BAAS}}               % Bulletin of the AAS
\def\jrasc{\aaref@jnl{JRASC}}             % Journal of the RAS of Canada
\def\memras{\aaref@jnl{MmRAS}}            % Memoirs of the RAS
\def\mnras{\aaref@jnl{MNRAS}}             % Monthly Notices of the RAS
\def\pra{\aaref@jnl{Phys.~Rev.~A}}        % Physical Review A: General Physics
\def\prb{\aaref@jnl{Phys.~Rev.~B}}        % Physical Review B: Solid State
\def\prc{\aaref@jnl{Phys.~Rev.~C}}        % Physical Review C
\def\prd{\aaref@jnl{Phys.~Rev.~D}}        % Physical Review D
\def\pre{\aaref@jnl{Phys.~Rev.~E}}        % Physical Review E
\def\prl{\aaref@jnl{Phys.~Rev.~Lett.}}    % Physical Review Letters
\def\pasp{\aaref@jnl{PASP}}               % Publications of the ASP
\def\pasj{\aaref@jnl{PASJ}}               % Publications of the ASJ
\def\qjras{\aaref@jnl{QJRAS}}             % Quarterly Journal of the RAS
\def\skytel{\aaref@jnl{S\&T}}             % Sky and Telescope
\def\solphys{\aaref@jnl{Sol.~Phys.}}      % Solar Physics
\def\sovast{\aaref@jnl{Soviet~Ast.}}      % Soviet Astronomy
\def\ssr{\aaref@jnl{Space~Sci.~Rev.}}     % Space Science Reviews
\def\zap{\aaref@jnl{ZAp}}                 % Zeitschrift fuer Astrophysik
\def\nat{\aaref@jnl{Nature}}              % Nature
\def\iaucirc{\aaref@jnl{IAU~Circ.}}       % IAU Cirulars
\def\aplett{\aaref@jnl{Astrophys.~Lett.}} % Astrophysics Letters
\def\apspr{\aaref@jnl{Astrophys.~Space~Phys.~Res.}}
                % Astrophysics Space Physics Research
\def\bain{\aaref@jnl{Bull.~Astron.~Inst.~Netherlands}} 
                % Bulletin Astronomical Institute of the Netherlands
\def\fcp{\aaref@jnl{Fund.~Cosmic~Phys.}}  % Fundamental Cosmic Physics
\def\gca{\aaref@jnl{Geochim.~Cosmochim.~Acta}}   % Geochimica Cosmochimica Acta
\def\grl{\aaref@jnl{Geophys.~Res.~Lett.}} % Geophysics Research Letters
\def\jcp{\aaref@jnl{J.~Chem.~Phys.}}      % Journal of Chemical Physics
\def\jgr{\aaref@jnl{J.~Geophys.~Res.}}    % Journal of Geophysics Research
\def\jqsrt{\aaref@jnl{J.~Quant.~Spec.~Radiat.~Transf.}}
                % Journal of Quantitiative Spectroscopy and Radiative Transfer
\def\memsai{\aaref@jnl{Mem.~Soc.~Astron.~Italiana}}
                % Mem. Societa Astronomica Italiana
\def\nphysa{\aaref@jnl{Nucl.~Phys.~A}}   % Nuclear Physics A
\def\physrep{\aaref@jnl{Phys.~Rep.}}   % Physics Reports
\def\physscr{\aaref@jnl{Phys.~Scr}}   % Physica Scripta
\def\planss{\aaref@jnl{Planet.~Space~Sci.}}   % Planetary Space Science
\def\procspie{\aaref@jnl{Proc.~SPIE}}   % Proceedings of the SPIE

\let\astap=\aap
\let\apjlett=\apjl
\let\apjsupp=\apjs
\let\applopt=\ao

\date{Accepted 2008 March ?. Received 2008 February ?; in original form 2008 February ?}

\pagerange{\pageref{firstpage}--\pageref{lastpage}} \pubyear{2008}

\maketitle

\label{firstpage}

\begin{abstract}
The origin of the magnetic fields in neutron stars, and the physical differences between magnetars and strongly magnetised radio pulsars are still under vigorous debate. It has been suggested that the properties of the progenitors of neutron stars (the massive OB stars), such as rotation, magnetic fields and mass, may play an important role in the outcome of core collapse leading to type II SNe.
Therefore, knowing the magnetic properties of the progenitor OB stars would be an important asset for constraining models of stellar evolution leading to the birth of a neutron star. 
We present here the beginning of a broad study with the goal of characterising the magnetic properties of main sequence massive OB stars. We report the detection of two new massive magnetic stars in the Orion Nebula Cluster: Par\,1772 (HD\,36982) and NU\,Ori (HD\,37061), for which the estimated dipole polar strengths, with $1\sigma$ error bars, are $1150^{+320}_{-200}$\,G and $650^{+220}_{-170}$\,G respectively. 

\end{abstract}


\section{Introduction}

Strong, organised magnetic fields are observed to exist in some main sequence stars of spectral type A, B and O.
Two general models have been proposed to explain the presence of these magnetic fields:
\begin{enumerate}
\item In the dynamo model, the field is generated by a dynamo mechanism, occurring classically in the convective regions or induced by strong shear during differential rotation.
\item In the fossil model, the field is a remnant from a dynamo active during a previous evolutionary phase, or swept up from the interstellar medium (ISM) during star formation. This scenario implies that the field must somehow survive the various internal structural changes encountered during stellar evolution. The magnetic flux is usually assumed to be conserved to some extent.
\end{enumerate}

Although dynamo models reproduce well the characteristics of late-type main sequence stars and giants, they fail to explain the fields of magnetic early-type stars, as their envelopes are primarily radiative. Some models of dynamo activity in the small convective cores of those stars have been put forward, but they still have fundamental difficulties reproducing the observed field characteristics \citep{2001ApJ...559.1094C}. Their simple magnetic geometries, lack of significant mass-field strength or period-field strength relation, and the fact that the observed characteristics of magnetic fields in pre-main sequence Herbig Ae/Be stars \citep{alecian08,2007A&A...462..293C,2007MNRAS.376..361F,2007MNRAS.376.1145W,2005A&A...442L..31W} are qualitatively identical to those of their main sequence descendants, point toward a fossil origin. Furthermore, the incidence, geometries and strengths of white dwarf magnetic fields are at least qualitatively compatible with evolution from magnetic main sequence A and B stars, suggesting that the fields of white dwarfs may also be of fossil origin \citep[e.g.][]{2005MNRAS.356.1576W}.

In more massive OB stars, magnetic fields have only been discovered recently, mostly via clues provided by unusual X-ray properties. Traditionally, the X-ray emission from O and B stars, with a typical level $L_X/L_{bol}\sim10^{-7}$, has been explained by radiative instabilities via a multitude of shocks in the wind \citep{1980ApJ...241..300L,1999ApJ...520..833O}. However, the very strong and rotationally modulated X-ray emission of the brightest Trapezium star, $\theta^1$\,Ori\,C (O7, P=15.4\,d, \citet{1997ApJ...478L..87G}) was explained by \citet{1997ApJ...485L..29B} in terms of the ``magnetically confined wind shock'' model (MWCS). In this model, the stellar magnetic field is sufficiently strong, and the radiative wind sufficiently weak, to allow a dipolar magnetic field to confine the outflowing wind in the immediate circumstellar environment, resulting in a closed magnetosphere with a large-scale equatorial shock which heats the wind plasma. In this way, the X-ray emission is enhanced and may be modulated by stellar rotation.
The MCWS model provided a quantitative prediction of a magnetic field
 in $\theta^1$\,Ori\,C; such a field ($1.1\pm0.1\rm\,kG$) was subsequently discovered by  \citet{2002MNRAS.333...55D}. At the present time, $\theta^1$\,Ori\,C and HD\,191612 \citep[1.5\,kG,][]{2006MNRAS.365L...6D} are the only known O-type stars with directly detected magnetic fields. However, it has been speculated that magnetism may be widespread among massive stars. Some clues to the presence of magnetic fields comes from X-ray photometry and spectroscopy \citep{2005ApJS..160..557S,2007ApJ...668..456W}, non-themal radio synchrotron emission \citep{2007A&A...470.1105S} and cyclical variations of UV wind spectral lines \citep{2003ASPC..305..333F,1996A&AS..116..257K}. Hence, this lack of magnetic field detection may well be due to the fact that direct measurement of magnetic fields present in the atmosphere of O-type and early-B type stars is extremely difficult. These difficulties arise from the small number of photospheric lines present in the optical spectrum and the large intrinsic width of the lines, worsened by the usual fast rotation of these stars.

Neutron stars, evolved from the massive OB-stars, are characterised by a wide range of magnetic field strengths. Inferred from spin down rates of radio pulsars, their strengths are in the range of $10^{11}$-$10^{14}$\,G. Two groups of neutron stars, the anomalous X-ray pulsars (AXPs) and the soft gamma repeaters (SGRs), host super-strong magnetic fields ($10^{14}$-$10^{15}$\,G), and are referred to as magnetars. It is thought that the physical distinction between radio pulsars and magnetars is not simply the dipole field strength, as there is a small population of radio pulsar {\bf with} fields at a magnetar-like level, but that does not show the same X-ray characteristics \citep{2005ApJ...618L..41K}. 
There is some observational evidence that neutron stars may evolve from stars as massive as $45\rm\,M_{\odot}$, and that many magnetars are linked strongly to these massive stars \citep{2005ApJ...620L..95G,2006ApJ...636L..41M}.

The magnetic flux of $\theta^1$\,Ori\,C ($45\rm\,M_{\odot}$) is  $(7\pm3)\times10^{27}\rm\,G\,cm^2$ (using the stellar radius from \citet{2006A&A...448..351S}). This magnetic flux is roughly of the same scale as the highest field magnetar SGR 1806-20\footnote{http://www.physics.mcgill.ca/$\sim$pulsar/magnetar/main.html.} ($\sim3\times10^{28}\rm\,G\,cm^2$, assuming a 10\,km radius). Therefore, in principle there is enough magnetic flux present in a massive magnetic star like $\theta^1$\,Ori\,C to explain the super-strong fields seen in some neutron stars, under the simple assumption that the magnetic flux is completely conserved during its post-MS stellar evolution and transformation into a neutron star. Furthermore, provided that OB-star fields are remnants from the ISM, the fossil hypothesis could provide a powerful explanation of the wide range of magnetic fields present in neutron stars \citep{2006MNRAS.367.1323F}. 

On the other hand, it has been suggested that neutron star magnetic fields could instead be generated during the core collapse itself, by a dynamo mechanism induced by differential rotation \citep{2006A&A...449..451B}. Present studies assume that any primordial fields present in the progenitor star are weak enough to be expelled by the dynamo process. However, if the initial field is strong enough, the evolution will be different, as this field is likely to interfere with differential rotation and therefore with the dynamo process itself \citep{2005ApJ...626..350H}.

Hence, there seem to be three fundamental parameters that may play key roles in the origin of neutron star magnetic fields, and in the explanation of the differences between magnetars and radio pulsars: the primordial magnetic field of the progenitor, the rotation of the star and its mass. Therefore, knowing the magnetic properties of the progenitor OB stars would be an important asset for constraining models of stellar evolution leading to neutron star birth. 
Many observational efforts are underway to characterise magnetic fields throughout massive star evolution. We present here the beginning of a broad study with the goal of characterising the magnetic properties of main sequence massive OB stars.

The Orion Nebula Cluster (ONC) presents a unique opportunity to characterise the magnetic fields of a nearby co-evolved and co-environmental population of massive OB stars. Furthermore, a {\it Chandra} large program, the {\it Chandra} Orion Ultradeep Project (COUP) was dedicated to observe the ONC in X-ray \citep{2005ApJS..160..557S}, enabling a study of the connections between stellar winds, magnetic fields and X-rays emission, which will be presented in a subsequent paper. The ONC contains 9 massive OB stars. They range from B3\,V ($\sim8\rm\,M_\odot$) to O7\,V ($\sim40\rm\,M_\odot$), approximately the mass range from which neutron stars are thought to be formed.
In this paper we report the detection of two new massive magnetic stars in the Orion Nebula Cluster.

\section{Observations}
\label{sec_obs}

\begin{figure*}  
\includegraphics[width=168mm]{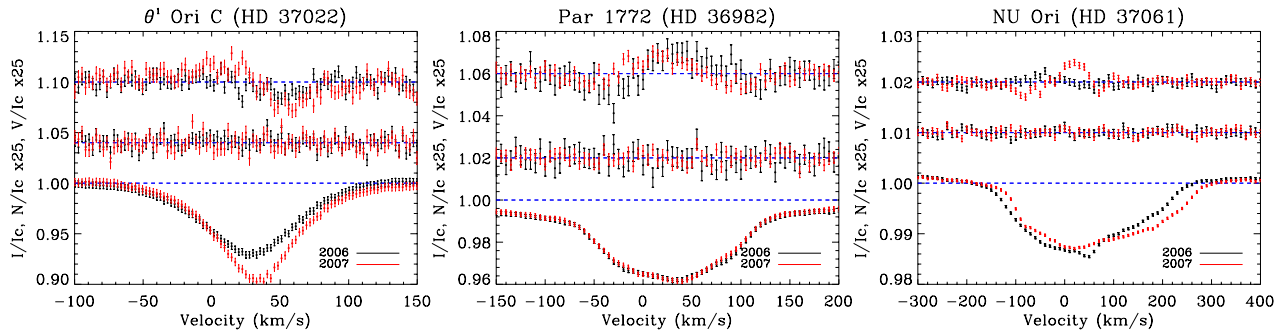}
\caption{\label{lsd} Least Squares Deconvolved profiles for $\theta^1$OriC (left), Par1772 (middle) and NU Ori (right). The curves are the mean Stokes I profiles (bottom), the mean Stokes V profiles (top) and the N diagnostic null profiles (middle), black for January 2006 and red for March 2007.} 
\end{figure*}

\begin{table*}
\centering
\begin{minipage}{170mm}
 \caption{\label{table}Observation log for the detected stars, along with detection diagnostics and derived longitudinal field components}
 \begin{tabular}{@{}llcccccrr}
  \hline
  Star 		&Date (UT)	& HJD 		& Total exp. time & Peak snr ${^{\rm a}}$ &LSD snr ${^{\rm a}}$& Detection &  P (\%) & B$_l (G)$  \\
  \hline
  $\theta^1$OriC 	& 2006-01-09	& 53744.792 	& 4\,800s & 1\,700 	& 3\,000 	& Marginal 	& $99.98$ 	& $131\pm56$ \\
  (O7V)				& 2007-03-10	& 54168.835 	& 3\,200s & 1\,600 	& 2\,600 	& Definite   	& $>99.99999$ 	& $471\pm53$\\
				& 2007-12-21	& 54456.748	& 3\,200s & 2\,000 	& 3\,700 	& None		& $66.8$		& $-53\pm44$\\
				& 2007-12-22	& 54457.748 	& 3\,200s & 1\,800 	& 3\,200 	& None		& $93.2$		& $122\pm50$\\
  Par 1772 		& 2006-01-12	& 53747.728 	& 9\,600s & 440 	& 2\,000 	& Definite 		& $99.99997$ 	& $-249\pm77$\\
  (B2V)				& 2007-03-07	& 54166.699 	& 9\,600s & 760 	& 3\,400 	& Definite   	& $>99.99999$		& $84\pm45$\\
				& 2007-11-11	&54416.550 	& 6\,000s & 370	& 1\,600 	& Marginal 	& $99.93$		& $-321\pm95$\\
  NU Ori 			& 2006-01-12	&53747.852 	& 9\,600s & 1\,300	& 14\,000 & None     		& $50.5$ 	& $82\pm52$\\
  (B0.5V)			& 2007-03-08	& 54167.703 	& 9\,600s & 1\,500 	& 15\,000 & Definite 		& $>99.99999$		& $-165\pm56$\\
  \hline
 \end{tabular}
 
\medskip
 ${^{\rm a}}$ Per 1.8\,km/s pixel for the summed spectra
 \end{minipage}
\end{table*}

We conducted spectropolarimetric observations with the ESPaDOnS spectropolarimeter at CFHT in January 2006 and March 2007. We obtained high-resolution (R$\sim$65,000) and high S/N spectra of 8 of the 9 massive OB stars of the ONC, in both epochs. Additional measurements of $\theta^1$\,Ori\,C and Par\,1772 were taken with ESPaDOnS in December 2007
and with ESPaDOnS's twin Narval, installed at TBL, France, in November 2007 respectively.

A complete circular polarisation observation consists of series of 4 sub-exposures between which the polarimeter quarter-wave plate is rotated back and forth between position angles, which make it possible to reduce systematic errors. 
For a complete description of observation procedures and reduction procedures with the Esprit reduction package (which is fundamentally the same as the Libre-Esprit package provided by CFHT), see \citet{1997MNRAS.291..658D}.

In order to increase the magnetic sensitivity of our data, we applied the Least Squares Deconvolution (LSD) procedure \citep{1997MNRAS.291..658D}, which enables the simultaneous use of many lines present in a spectrum to detect a magnetic field Stokes V signature. The line masks were carefully chosen to exclude Balmer lines and any other lines blended with them.
Using the LSD technique, we found clear Stokes V signatures for 3 stars: the previously-detected $\theta^1$\,Ori\,C, as well as Par\,1772 (HD\,36982) and NU Ori (HD\,37061). The LSD profiles for these three stars are shown in Figure \ref{lsd}.
We used the statistical test described by \citet{1997MNRAS.291..658D} to diagnose the presence of a signal in either mean Stokes V or diagnostic null profiles. A signal is unambiguously detected inside the spectral line range whenever the associated detection probability is larger than 99.999 per cent (corresponding to a false alarm probability smaller than $10^{-5}$). The associated longitudinal field can be estimated with the Stokes I and V profiles \citep[e.g.][]{2000MNRAS.313..851W}. It is important to note that the longitudinal field is only given as an indicator of the field strength and not the main diagnostic for the presence of a magnetic field, because a magnetic configuration with a null longitudinal component can still produce a non-null Stokes V signature. Table \ref{table} presents an observing log for the detected stars, along with the detection probability inside the spectral line range and the derived longitudinal field component.

\begin{figure} 
\includegraphics[width=75mm]{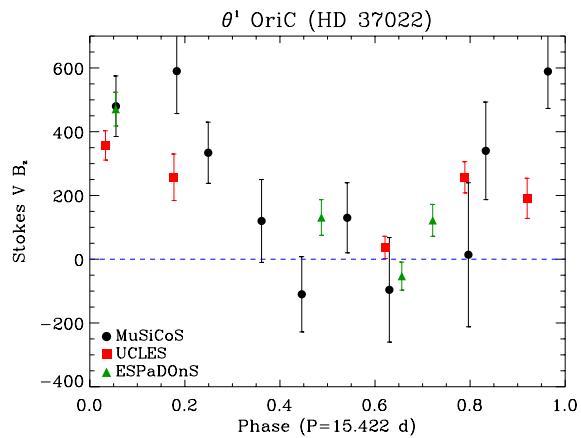}
\caption{\label{fig_long} Longitudinal magnetic field measurements for $\theta^1$\,Ori\,C. The MuSiCoS measurements (back circles) are from \citet{2006A&A...451..195W}, the UCLES measurement (red squares) are from \citet{2002MNRAS.333...55D} and the ESPaDOnS measurements (green triangles) are from this work.} 
\end{figure}

{\bf $\theta^1$\,Ori\,C} is the canonical example of a magnetic O star showing rotationally-modulated spectral and X-ray variations caused by magnetic confinement of its stellar wind \citep{1997ApJ...485L..29B,2002MNRAS.333...55D}. This most massive star of the Trapezium is a speckle binary with a $0^"_.037$ separation, composed of a 45\,M$_\odot$ primary (O7V) and a $\ga 6$\,M$_\odot$ secondary \citep{2003A&A...402..267S}.
Optical, UV and X-ray features all vary according to a $15.422\pm0.002$\,d period \citep{1997ApJ...478L..87G,1996A&A...312..539S}. We obtained 4 observations of this star, and according to this ephemeris, our observations correspond to phase 0.49, 0.05, 0.66 and 0.72 for January 2006, March 2007 and the 2 observations of December 2007 respectively. 
The corresponding derived longitudinal fields are in good agreement with previous spectropolarimetric measurements \citep{2002MNRAS.333...55D,2006A&A...451..195W}, and therefore with their derived dipolar field strength of about 1.1\,kG (Fig. \ref{fig_long}).

{\bf Par\,1772} is a main sequence (or possibly pre-main sequence) B2 star ($\sim6\rm\,M_\odot$), with a projected rotational velocity of $80\pm20\rm\,km/s$ \citep{2004ApJ...601..979W}. The March 2007 Stokes V signature is a good example of a cross-over signature, where the longitudinal field component is nearly null (here $84\pm45$\,G), but the symmetry of the polarized Zeeman components is broken by Doppler shifts induced by stellar rotation. 
The measurement of  $91\pm193$\,G obtained by \citet{2006A&A...450..777B} with FORS1 at VLT was likely at such a cross-over phase, the magnetic field going undetected because of the lower spectral resolving power of that instrument.

{\bf NU Ori} is a triple system, containing a B0.5V primary ($14\rm\,M_\odot$), along with a spectroscopic companion of $\sim3\rm\,M_\odot$ (component C, 80 mas separation) and a $\sim1\rm\,M_\odot$ visual companion (component B) with a $471\pm17\rm\,\,mas$ separation \citep{1999NewA....4..531P}. The primary is a rapidly rotating star, with a $v\sin i=225\pm50\rm\,km/s$ \citep{2004ApJ...601..979W}, making it the most rapidly rotating early-B star with a detected field. 
Although such high rotational velocity usually occurs only in Be stars, the small and narrow emission in H$\alpha$, $\beta$ and $\gamma$ seems more related to nebular emission than to a Be behaviour. 
While there was no formal signal detection for the January 2006 observation, the March 2007 observation showed a definite signal detection. 
A close inspection of the intensity spectrum revealed the weak, sharp spectral lines of the spectroscopic companion. The width of the Stokes V signatures compared to the width of the companion spectral lines rules out associating the polarization signature with the companion -- the magnetic field and the profile asymmetries are clearly intrinsic to the primary.

\section{Surface magnetic fields}

\begin{figure} 
\includegraphics[width=75mm]{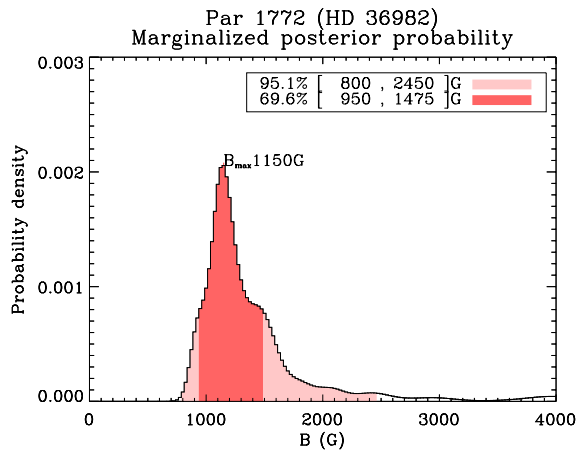}
\caption{\label{par1772_bayes} Marginalized probability densities of the dipole polar field strength for Par1772. The magnetic field strength 95\% credible region is filled in light color. The 68.3\% credible region used to calculate the 1$\sigma$ error bars is filled in dark color.} 
\end{figure}

\begin{figure} 
\includegraphics[width=75mm]{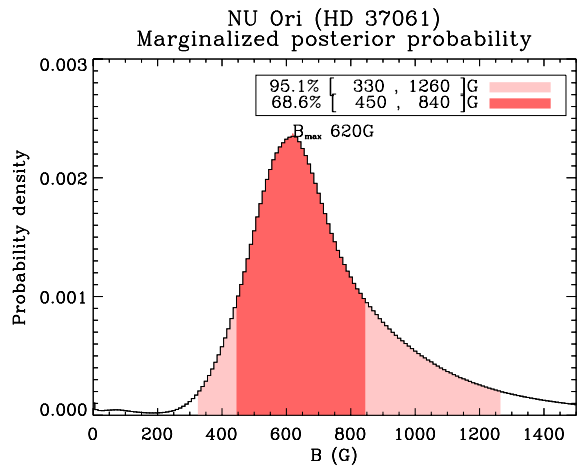}
\caption{\label{nuori_bayes}The same as Figure \ref{par1772_bayes} for NU Ori.} 
\end{figure}

In order to extract the surface field characteristics from the observed Stokes V profiles, we compared them with theoretical profiles for a large grid of dipolar magnetic field configurations, calculated with the polarised LTE radiative transfer code Zeeman2 \citep{1988ApJ...326..967L,2001A&A...374..265W}. 
We sampled the 4-dimensional parameter space ($i$, $\beta$, $\varphi$, B) which describes a centered dipolar magnetic configuration. In such a model, $i$ is the projected inclination of the rotation axis to the line of sight, $\beta$ is the obliquity of the magnetic axis with respect to the rotation axis, $\varphi$ is the rotational phase and B is the polar field strength at the stellar surface. 
For each configuration, we calculated the reduced $\chi^2$ of the model fit to the observed mean Stokes V profiles. Assuming that only the phase may change between two observations of a given star, the goodness-of-fit of a given rotation-independent ($i$, $\beta$, B)-configuration is expressed in terms of Bayesian probability density. This ensures that a good magnetic ($i$, $\beta$, B)-configuration can produce Stokes V profiles that fit all observations, as the rotational period is not known with enough accuracy to determine {\it a priori} the phase difference. Any features that cannot be explained by the rotating dipole model are treated formally as additional Gaussian noise, which will lead to the most conservative estimates of the parameters, according to the maximum entropy principle. 

We can determine the probability density of the field strength by marginalizing over inclination and obliquity. We then extract a 95\% credible region for the surface dipole field strength of each star with the technique described by \citet{2005blda.book.....G}. Figure \ref{par1772_bayes} and Figure \ref{nuori_bayes} show the resulting probability density functions for Par\,1772 and NU\,Ori respectively. The 95\% credible regions are [800, 2450]\,G for Par\,1772 and [370, 1220]\,G for NU\,Ori. The inferred values of the dipole polar strength, with $1\sigma$ error bars, are then $1150^{+320}_{-200}$\,G and $650^{+220}_{-170}$\,G respectively.

\section{Discussion}

As an illustrative example of the potential of these new data to constrain models of neutron star field origins, we can compare them with the predictions of \citet{2006MNRAS.367.1323F} of {\bf the} magnetic field distribution of massive stars (8-45\,M$_\odot$) on the main sequence. This distribution is based on the observed properties of radio pulsars from the 1374-MHz Parkes Multi-Beam Survey of isolated radio pulsar, assuming a simple fossil field origin with a complete conservation of magnetic flux.  
They obtained a continuous magnetic field distribution in the progenitor OB stars, peaking at $\sim46\rm\,G$ with 5 per cent\footnote{Although \citet{2006MNRAS.367.1323F} state 8\%, recalculation based on the detailed model distribution provided by L. Ferrario gives 5\%} of the stars having a field in excess of 1\,kG.

Of course, our sample contains only 8 stars, but we can still make some rough comparisons. 
Taking the predicted field strength distribution, we assume that it is the true parent distribution from which we draw a random sample of 8 stars. We define three possible outcomes: [0-500]\,G, [500-1000]\,G and over 1000\,G, with respective probabilities derived from the parent theoretical distribution. According to the multinomial distribution,  the probability of obtaining the distribution of magnetic field strengths observed in the ONC is about 1\%. 

This result might be interpreted, at first glance, to suggest that massive OB stars are more magnetic than it would be required to explain the magnetic fields of radio pulsars. 
However, some points are important to consider: 
(i) The sample of observed stars may not be representative of a general parent distribution, as the stars all come from the same cluster. This region could be unusually magnetic, especially if the fields of the OB stars themselves are also of fossil origin from the ISM.
(ii) The assumed parent distribution is not in fact the true parent distribution because some assumptions are incorrect, or some elements are missing from the model. Examples of such missing physics might be partial flux conservation or the influence of dynamo processes during core collapse.

In order to better explore these possibilities, a larger sample of OB stars, from clusters and from the field, must be studied in order to increase the population of neutron star progenitors with known magnetic properties. Our team has undertaken an extensive spectropolarimetric study of massive stars in other young star clusters to provide these important data.

\section*{Acknowledgments}
  VP acknowledges support from Fonds qu\'eb\'ecois de la recherche sur la nature et les technologies.
  LD acknowledges support from  the Canada Research Chair program and the Discovery Grants programme of the Natural Science and Engineering Research Council of Canada.
  GAW  acknowledges support from the Discovery Grants programme of the Natural Science and Engineering Research Council of Canada.
  EA is supported by the Marie Curie FP6 program.
Finally, we thank the anonymous referee, whose helpful comments led to the improvement of the paper.

\bibliographystyle{mn2e}
\bibliography{art_onc}

\begin{thebibliography}{}

\bibitem[\protect\citeauthoryear{{Alecian}, {Catala}, {Wade}, {Donati} {Petit},
  J.-F., {Landstreet}, {B\"om}, {Bouret}, {Bagnulo}, {Folsom}, {Grunhut} \&
  {Silvester}}{{Alecian} et~al.}{2008}]{alecian08}
{Alecian} E.,  {Catala} C.,  {Wade} G.,  {Donati} {Petit} P.,  J.-F.
  {Landstreet} J.,  {B\"om} T.,  {Bouret} J.-C.,  {Bagnulo} S.,  {Folsom} C.,
  {Grunhut} J.,    {Silvester} J.,  2008, MNRAS, p. accepted

\bibitem[\protect\citeauthoryear{{Babel} \& {Montmerle}}{{Babel} \&
  {Montmerle}}{1997}]{1997ApJ...485L..29B}
{Babel} J.,  {Montmerle} T.,  1997, \apjl, 485, L29

\bibitem[\protect\citeauthoryear{{Bagnulo}, {Landstreet}, {Mason}, {Andretta},
  {Silaj} \& {Wade}}{{Bagnulo} et~al.}{2006}]{2006A&A...450..777B}
{Bagnulo} S.,  {Landstreet} J.~D.,  {Mason} E.,  {Andretta} V.,  {Silaj} J.,
  {Wade} G.~A.,  2006, \aap, 450, 777

\bibitem[\protect\citeauthoryear{{Braithwaite}}{{Braithwaite}}{2006}]{2006A&A.%
..449..451B}
{Braithwaite} J.,  2006, \aap, 449, 451

\bibitem[\protect\citeauthoryear{{Catala}, {Alecian}, {Donati}, {Wade},
  {Landstreet}, {B\"ohm}, {Bouret}, {Bagnulo}, {Folsom} \&
  {Silvester}}{{Catala} et~al.}{2007}]{2007A&A...462..293C}
{Catala} C.,  {Alecian} E.,  {Donati} J.-F.,  {Wade} G.~A.,  {Landstreet}
  J.~D.,  {B\"ohm} T.,  {Bouret} J.-C.,  {Bagnulo} S.,  {Folsom} C.,
  {Silvester} J.,  2007, \aap, 462, 293

\bibitem[\protect\citeauthoryear{{Charbonneau} \& {MacGregor}}{{Charbonneau} \&
  {MacGregor}}{2001}]{2001ApJ...559.1094C}
{Charbonneau} P.,  {MacGregor} K.~B.,  2001, \apj, 559, 1094

\bibitem[\protect\citeauthoryear{{Donati}, {Babel}, {Harries}, {Howarth},
  {Petit} \& {Semel}}{{Donati} et~al.}{2002}]{2002MNRAS.333...55D}
{Donati} J.-F.,  {Babel} J.,  {Harries} T.~J.,  {Howarth} I.~D.,  {Petit} P.,
   {Semel} M.,  2002, \mnras, 333, 55

\bibitem[\protect\citeauthoryear{{Donati}, {Howarth}, {Bouret}, {Petit},
  {Catala} \& {Landstreet}}{{Donati} et~al.}{2006}]{2006MNRAS.365L...6D}
{Donati} J.-F.,  {Howarth} I.~D.,  {Bouret} J.-C.,  {Petit} P.,  {Catala} C.,
   {Landstreet} J.,  2006, \mnras, 365, L6

\bibitem[\protect\citeauthoryear{{Donati}, {Semel}, {Carter}, {Rees} \&
  {Collier Cameron}}{{Donati} et~al.}{1997}]{1997MNRAS.291..658D}
{Donati} J.-F.,  {Semel} M.,  {Carter} B.~D.,  {Rees} D.~E.,    {Collier
  Cameron} A.,  1997, \mnras, 291, 658

\bibitem[\protect\citeauthoryear{{Ferrario} \& {Wickramasinghe}}{{Ferrario} \&
  {Wickramasinghe}}{2006}]{2006MNRAS.367.1323F}
{Ferrario} L.,  {Wickramasinghe} D.,  2006, \mnras, 367, 1323

\bibitem[\protect\citeauthoryear{{Folsom}, {Wade}, {Bagnulo} \&
  {Landstreet}}{{Folsom} et~al.}{2007}]{2007MNRAS.376..361F}
{Folsom} C.~P.,  {Wade} G.~A.,  {Bagnulo} S.,    {Landstreet} J.~D.,  2007,
  \mnras, 376, 361

\bibitem[\protect\citeauthoryear{{Fullerton}}{{Fullerton}}{2003}]{2003ASPC..30%
5..333F}
{Fullerton} A.~W.,  2003, in {Balona} L.~A.,  {Henrichs} H.~F.,   {Medupe} R.,
  eds, Magnetic Fields in O, B and A Stars, Vol.~305 of ASPC Series.
p.~333

\bibitem[\protect\citeauthoryear{{Gaensler}, {McClure-Griffiths}, {Oey},
  {Haverkorn}, {Dickey} \& {Green}}{{Gaensler}
  et~al.}{2005}]{2005ApJ...620L..95G}
{Gaensler} B.~M.,  {McClure-Griffiths} N.~M.,  {Oey} M.~S.,  {Haverkorn} M.,
  {Dickey} J.~M.,    {Green} A.~J.,  2005, \apjl, 620, L95

\bibitem[\protect\citeauthoryear{{Gagn\'e}, {Caillault}, {Stauffer} \&
  {Linsky}}{{Gagn\'e} et~al.}{1997}]{1997ApJ...478L..87G}
{Gagn\'e} M.,  {Caillault} J.-P.,  {Stauffer} J.~R.,    {Linsky} J.~L.,  1997,
  \apjl, 478, L87

\bibitem[\protect\citeauthoryear{{Gregory}}{{Gregory}}{2005}]{2005blda.book...%
..G}
{Gregory} P.~C.,  2005, {Bayesian Logical Data Analysis for the Physical
  Sciences}.
Edited by P.~C.~Gregory.~;~Published by Cambridge University Press, Cambridge,
  UK.

\bibitem[\protect\citeauthoryear{{Heger}, {Woosley} \& {Spruit}}{{Heger}
  et~al.}{2005}]{2005ApJ...626..350H}
{Heger} A.,  {Woosley} S.~E.,    {Spruit} H.~C.,  2005, \apj, 626, 350

\bibitem[\protect\citeauthoryear{{Kaper}, {Henrichs}, {Nichols}, {Snoek},
  {Volten} \& {Zwarthoed}}{{Kaper} et~al.}{1996}]{1996A&AS..116..257K}
{Kaper} L.,  {Henrichs} H.~F.,  {Nichols} J.~S.,  {Snoek} L.~C.,  {Volten} H.,
    {Zwarthoed} G.~A.~A.,  1996, \aaps, 116, 257

\bibitem[\protect\citeauthoryear{{Kaspi} \& {McLaughlin}}{{Kaspi} \&
  {McLaughlin}}{2005}]{2005ApJ...618L..41K}
{Kaspi} V.~M.,  {McLaughlin} M.~A.,  2005, \apjl, 618, L41

\bibitem[\protect\citeauthoryear{{Landstreet}}{{Landstreet}}{1988}]{1988ApJ...%
326..967L}
{Landstreet} J.~D.,  1988, \apj, 326, 967

\bibitem[\protect\citeauthoryear{{Lucy} \& {White}}{{Lucy} \&
  {White}}{1980}]{1980ApJ...241..300L}
{Lucy} L.~B.,  {White} R.~L.,  1980, \apj, 241, 300

\bibitem[\protect\citeauthoryear{{Muno}, {Clark}, {Crowther}, {Dougherty}, {de
  Grijs}, {Law}, {McMillan}, {Morris}, {Negueruela}, {Pooley}, {Portegies
  Zwart} \& {Yusef-Zadeh}}{{Muno} et~al.}{2006}]{2006ApJ...636L..41M}
{Muno} M.~P.,  {Clark} J.~S.,  {Crowther} P.~A.,  {Dougherty} S.~M.,  {de
  Grijs} R.,  {Law} C.,  {McMillan} S.~L.~W.,  {Morris} M.~R.,  {Negueruela}
  I.,  {Pooley} D.,  {Portegies Zwart} S.,    {Yusef-Zadeh} F.,  2006, \apjl,
  636, L41

\bibitem[\protect\citeauthoryear{{Owocki} \& {Cohen}}{{Owocki} \&
  {Cohen}}{1999}]{1999ApJ...520..833O}
{Owocki} S.~P.,  {Cohen} D.~H.,  1999, \apj, 520, 833

\bibitem[\protect\citeauthoryear{{Preibisch}, {Balega}, {Hofmann}, {Weigelt} \&
  {Zinnecker}}{{Preibisch} et~al.}{1999}]{1999NewA....4..531P}
{Preibisch} T.,  {Balega} Y.,  {Hofmann} K.-H.,  {Weigelt} G.,    {Zinnecker}
  H.,  1999, New Astronomy, 4, 531

\bibitem[\protect\citeauthoryear{{Schertl}, {Balega}, {Preibisch} \&
  {Weigelt}}{{Schertl} et~al.}{2003}]{2003A&A...402..267S}
{Schertl} D.,  {Balega} Y.~Y.,  {Preibisch} T.,    {Weigelt} G.,  2003, \aap,
  402, 267

\bibitem[\protect\citeauthoryear{{Schnerr}, {Rygl}, {van der Horst},
  {Oosterloo}, {Miller-Jones}, {Henrichs}, {Spoelstra} \& {Foley}}{{Schnerr}
  et~al.}{2007}]{2007A&A...470.1105S}
{Schnerr} R.~S.,  {Rygl} K.~L.~J.,  {van der Horst} A.~J.,  {Oosterloo} T.~A.,
  {Miller-Jones} J.~C.~A.,  {Henrichs} H.~F.,  {Spoelstra} T.~A.~T.,    {Foley}
  A.~R.,  2007, \aap, 470, 1105

\bibitem[\protect\citeauthoryear{{Sim\'on-D{\'{\i}}az}, {Herrero}, {Esteban} \&
  {Najarro}}{{Sim\'on-D{\'{\i}}az} et~al.}{2006}]{2006A&A...448..351S}
{Sim\'on-D{\'{\i}}az} S.,  {Herrero} A.,  {Esteban} C.,    {Najarro} F.,  2006,
  \aap, 448, 351

\bibitem[\protect\citeauthoryear{{Stahl}, {Kaufer}, {Rivinius}, {Szeifert},
  {Wolf}, {Gaeng}, {Gummersbach}, {Jankovics}, {Kovacs}, {Mandel}, {Pakull} \&
  {Peitz}}{{Stahl} et~al.}{1996}]{1996A&A...312..539S}
{Stahl} O.,  {Kaufer} A.,  {Rivinius} T.,  {Szeifert} T.,  {Wolf} B.,  {Gaeng}
  T.,  {Gummersbach} C.~A.,  {Jankovics} I.,  {Kovacs} J.,  {Mandel} H.,
  {Pakull} M.~W.,    {Peitz} J.,  1996, \aap, 312, 539

\bibitem[\protect\citeauthoryear{{Stelzer}, {Flaccomio}, {Montmerle}, {Micela},
  {Sciortino}, {Favata}, {Preibisch} \& {Feigelson}}{{Stelzer}
  et~al.}{2005}]{2005ApJS..160..557S}
{Stelzer} B.,  {Flaccomio} E.,  {Montmerle} T.,  {Micela} G.,  {Sciortino} S.,
  {Favata} F.,  {Preibisch} T.,    {Feigelson} E.~D.,  2005, \apjs, 160, 557

\bibitem[\protect\citeauthoryear{{Wade}, {Bagnulo}, {Drouin}, {Landstreet} \&
  {Monin}}{{Wade} et~al.}{2007}]{2007MNRAS.376.1145W}
{Wade} G.~A.,  {Bagnulo} S.,  {Drouin} D.,  {Landstreet} J.~D.,    {Monin} D.,
  2007, \mnras, 376, 1145

\bibitem[\protect\citeauthoryear{{Wade}, {Bagnulo}, {Kochukhov}, {Landstreet},
  {Piskunov} \& {Stift}}{{Wade} et~al.}{2001}]{2001A&A...374..265W}
{Wade} G.~A.,  {Bagnulo} S.,  {Kochukhov} O.,  {Landstreet} J.~D.,  {Piskunov}
  N.,    {Stift} M.~J.,  2001, \aap, 374, 265

\bibitem[\protect\citeauthoryear{{Wade}, {Donati}, {Landstreet} \&
  {Shorlin}}{{Wade} et~al.}{2000}]{2000MNRAS.313..851W}
{Wade} G.~A.,  {Donati} J.-F.,  {Landstreet} J.~D.,    {Shorlin} S.~L.~S.,
  2000, \mnras, 313, 851

\bibitem[\protect\citeauthoryear{{Wade}, {Drouin}, {Bagnulo}, {Landstreet},
  {Mason}, {Silvester}, {Alecian}, {B\"ohm}, {Bouret}, {Catala} \&
  {Donati}}{{Wade} et~al.}{2005}]{2005A&A...442L..31W}
{Wade} G.~A.,  {Drouin} D.,  {Bagnulo} S.,  {Landstreet} J.~D.,  {Mason} E.,
  {Silvester} J.,  {Alecian} E.,  {B\"ohm} T.,  {Bouret} J.-C.,  {Catala} C.,
   {Donati} J.-F.,  2005, \aap, 442, L31

\bibitem[\protect\citeauthoryear{{Wade}, {Fullerton}, {Donati}, {Landstreet},
  {Petit} \& {Strasser}}{{Wade} et~al.}{2006}]{2006A&A...451..195W}
{Wade} G.~A.,  {Fullerton} A.~W.,  {Donati} J.-F.,  {Landstreet} J.~D.,
  {Petit} P.,    {Strasser} S.,  2006, \aap, 451, 195

\bibitem[\protect\citeauthoryear{{Waldron} \& {Cassinelli}}{{Waldron} \&
  {Cassinelli}}{2007}]{2007ApJ...668..456W}
{Waldron} W.~L.,  {Cassinelli} J.~P.,  2007, \apj, 668, 456

\bibitem[\protect\citeauthoryear{{Wickramasinghe} \&
  {Ferrario}}{{Wickramasinghe} \& {Ferrario}}{2005}]{2005MNRAS.356.1576W}
{Wickramasinghe} D.~T.,  {Ferrario} L.,  2005, \mnras, 356, 1576

\bibitem[\protect\citeauthoryear{{Wolff}, {Strom} \& {Hillenbrand}}{{Wolff}
  et~al.}{2004}]{2004ApJ...601..979W}
{Wolff} S.~C.,  {Strom} S.~E.,    {Hillenbrand} L.~A.,  2004, \apj, 601, 979

\end{thebibliography}
\clearpage

\label{lastpage}

\end{document}